\documentstyle{mn}
\begin{document}
\def\simlt{\mathrel{\rlap{\lower 3pt\hbox{$\sim$}} \raise
        2.0pt\hbox{$<$}}} \def\simgt{\mathrel{\rlap{\lower
        3pt\hbox{$\sim$}} \raise 2.0pt\hbox{$>$}}}
\title[The Halo Distribution of 2dF Galaxies]
{The Halo Distribution of 2dF Galaxies }
\author[M.Magliocchetti, C.Porciani]
{Manuela Magliocchetti$^{1}$ \& Cristiano Porciani$^{2,3}$\\ 
$^{1}$SISSA, Via Beirut 4, 34014, Trieste, Italy\\ 
$^{2}$ Institute of Astronomy, Madingley Road, CB30HA, Cambridge, UK\\
$^{3}$ Institut f\"ur Astronomie, HPF G3.2, ETH H\"onggerberg, 8093 Z\"urich,
Switzerland}
\maketitle
\vspace {7cm}

\begin{abstract}
We use the clustering results obtained by Madgwick et al. (2003) for a sample 
of 96,791 2dF galaxies with redshift $0.01 < z < 0.15$ to study the 
distribution of late-type and early-type galaxies within dark matter haloes 
of different mass. Within the framework of our models,
galaxies of both classes are found to be as spatially 
concentrated as the dark matter within haloes even though, while the 
distribution of star-forming galaxies can also allow for some steeper 
profiles, this is drastically ruled out in the case of early-type 
galaxies. We also find evidence for morphological segregation, as  
late-type galaxies appear to be distributed within haloes of mass scales 
corresponding to groups and clusters up to about two virial radii, while 
passive objects show a preference to reside closer to the halo centre. 
If we assume a broken power-law of the form $\langle N_{\rm gal}\rangle(m)=(m/m_0)^
{\alpha_1}$ for $m_{\rm cut}\le m<m_0$ and $\langle N_{\rm gal}\rangle(m)=(m/m_0)
^{\alpha_2}$ at higher masses to describe the dependence of  
the average number of galaxies within haloes on the halo mass, fits to the 
data show that star-forming galaxies start appearing in haloes of masses 
$m_{\rm cut}\simeq 10^{11}m_\odot$, much smaller than what is obtained for 
early-type galaxies ($m_{\rm cut}\simeq 10^{12.6}m_\odot$). In the high-mass regime 
$m\ge m_0$, $\langle N_{\rm gal}\rangle$ increases with halo mass more slowly 
($\alpha_2\simeq 0.7$) 
in the case of late-type galaxies than for passive objects which present 
$\alpha_2\simeq 1.1$. The above results imply that late-type galaxies dominate the 2dF 
counts at all mass scales.
We stress that -- at variance with previous statements -- there is no 
degeneracy in the determination of the best functional 
forms for $\rho(r)$ and $\langle N_{\rm gal}\rangle$, as they affect the 
behaviour 
of the galaxy-galaxy correlation function on different scales.  
\end{abstract}


\section{Introduction}
Measurements of the galaxy-galaxy correlation function contain a wealth
of information both on the underlying cosmological model 
and on the physical processes connected with the formation and evolution of 
galaxies. 
Untangling these two effects from the observed clustering signal of a 
particular class of sources is, however, not an easy task.
For instance,   
the physics of galaxy formation affects the 
relationship between the distribution of luminous and dark matter 
(the so-called ``bias'') so that different types 
of galaxies are expected to exhibit different clustering properties. 
This has indeed been observed (see e.g. Loveday et al., 1995; Guzzo et al., 
1997; Loveday, Tresse \& Maddox, 1999; Magliocchetti et al., 2000 just to 
mention few) in the past decade, when large-area surveys started 
including enough sources to allow for precision clustering statistics.

From a theoretical point of view, the relationship between dark matter and 
galaxy distribution has not yet fully been understood since, while the 
dynamics of dark matter is only driven by gravity and fully determined by 
the choice of an appropriate cosmological model (see e.g. Jenkins et al., 
1998), the situation gets increasingly more difficult as one tries to 
model the physical processes playing a role in the process of galaxy 
formation.\\ 
As a first approximation, galaxies can be associated with 
the dark matter haloes in which they reside (in a one-to-one 
relationship), so that their clustering properties can be derived  
within the framework of the halo-model developed by Mo \& White (1996). 
Such models have been proved extremely useful to describe the clustering 
of high-redshift sources such as quasars, Lyman Break and SCUBA 
galaxies (see e.g. 
Matarrese et al., 1997; Moscardini et al., 1998; 
Martini \& Weinberg, 2001; Magliocchetti et al., 2001; 
Porciani \& Giavalisco, 2002) where the assumption of one 
such object per halo 
can be considered a reasonable guess.\\
The validity of this approach however breaks down as one moves to 
objects with higher number densities such as low-redshift galaxies. 
The distribution of these kind of sources within dark matter haloes is in 
general an unknown quantity which will  
depend on the efficiency of galaxy formation via some complicated physics 
connected to processes such as gas cooling and/or supernova feedback (see e.g. 
Somerville et al., 2001; Benson et al., 2001).
 
The analytical connection between the distribution of sources within dark 
matter haloes 
and their clustering properties has been studied in detail by a number of 
recent papers (Peacock \& Smith, 2000; Seljak, 2000; Scoccimarro et al., 2001; 
Bullock, Wechsler \& Somerville 2002;
Marinoni \& Hudson, 2002; Berlind \& Weinberg, 2001; 
Moustakas \& Somerville, 2002; Yang, Mo \& van den Bosch 2003;
van den Bosch, Yang \& Mo, 2003; see also 
Cooray \& Sheth, 2002 for an extensive review on the topic)
which mainly focus 
on the issue of the halo occupation function i.e. 
the probability distribution of the 
number of galaxies brighter than some luminosity threshold hosted 
by a virialized halo of given mass. 
Within this framework, the distribution of galaxies within haloes 
is shown to determine galaxy-galaxy clustering on small scales, being 
responsible for the observed power-law behaviour at separations
$r\simlt 3$~Mpc. 

A number of parameters are necessary to describe the halo occupation   
distribution (i.e. the probability of finding $N_{\rm gal}$ 
galaxies in a halo of mass 
$m$) of a class of galaxies. These parameters -- expected to vary with galaxy 
type -- cannot be worked out from 
first principles and have to rely for their determination either on  
comparisons with results from semi-analytical models or on statistical 
measurements coming from large data-sets, with an obvious preference for this 
second approach.\\
The two last-generation 2dF and SDSS Galaxy Redshift Surveys (Colless et al., 
2001; York et al., 2000) come in our help since -- with their unprecedented 
precision in measuring galaxy clustering -- they can in principle constrain 
the functional form of the halo occupation distribution (see also Zehavi et al., 2003).

Lower-order clustering measurements such as the spatial two-point correlation 
function are already available for samples of 
2dF sources (Norberg et al., 2001; Hawkins et al., 2003; Madgwick et al., 2003). 
In this work we will use the results of Madgwick et al. (2003) on the 
clustering properties of late-type and early-type galaxies to investigate 
possible differences in the processes responsible for the birth and 
evolution of these two classes of sources.       

The layout of the paper is as follows: Section 2 introduces the formalism 
necessary to describe galaxy clustering within the halo 
occupation distribution context. Key ingredients for this kind of analysis 
are the average number $\langle N_{\rm gal}\rangle(m)$ of galaxies hosted in dark matter 
haloes of mass $m$, a measure of 
the spread $\langle N_{\rm gal}(N_{\rm gal}-1)\rangle(m)$ 
about this mean value 
and the spatial distribution $\rho(r)$ of galaxies within their haloes.
In Section 3 we briefly describe the 2dF Galaxy Redshift Survey with 
particular attention devoted to measurements of the luminosity function and 
correlation function for early-type (i.e. passively evolving) and 
late-type (which are still in the process of active star formation) galaxies,
and derive estimates for the number density of these sources. Section 4 
presents and discusses our results on the distribution of different types of 
galaxies within dark matter haloes as obtained by comparing predictions 
on their number density and correlation function  
with 2dF observations, while Section 5 summarizes our conclusions.

Unless differently stated, throughout this work we will assume 
that the density parameter $\Omega_0=0.3$, the vacuum energy density 
$\Lambda=0.7$, 
the present-day value of the Hubble parameter in units of $100$ km/s/Mpc
$h_0=0.65$ and 
$\sigma_8=0.8$ (with $\sigma_8$ the rms linear density fluctuation within a 
sphere with a radius of $8 h^{-1}$ Mpc), as the latest results 
from the joint analysis of CMB and 2dF data seem to indicate 
(see e.g. Lahav et al., 2002; Spergel et al., 2003).
Note, however, that the normalization of the linear power spectrum of density
fluctuations is still very controversial:
estimates of $\sigma_8$ from either weak-lensing or cluster abundances
range between 0.6 and 1.0, 
while some analyses of galaxy clustering seem
to favour values $\sim 0.7$ (van den Bosch, Mo \& Yang 2003).
Our results will slightly depend on the assumed value for $\sigma_8$.

\section{Galaxy Clustering: The Theory}
The purpose of this Section is to introduce the formalism and specify the 
 ingredients necessary to describe galaxy clustering at the 2-point level.
Our approach follows the one adopted by Scoccimarro et al. (2001) which is in 
 turn based on the analysis performed by Scherrer \& Bertschinger (1991).
In this framework, the galaxy-galaxy correlation function can be written as \\
\begin{eqnarray}
\xi_g({\bf x}-{\bf x}^\prime)=
\xi_g^{1h}({\bf x}-
{\bf x}^\prime)+\xi_g^{2h}({\bf x}-{\bf x}^\prime),
\label{eq:xi}
\end{eqnarray}
with
\begin{eqnarray}
\xi_g^{1h}=\frac{1}{\bar{n}_g^2 }
\int n(m) \langle N_{\rm gal}(N_{\rm gal}-1)\rangle(m) dm \times\nonumber \\
\int \rho_m({\bf y})\rho_m({\bf y}+{\bf x}-{\bf x}^\prime) d^3y,
\end{eqnarray}
and
\begin{eqnarray}
\xi_g^{2h}=\frac{1}{\bar{n}_g^2 }
\int n(m_1) \langle N_{\rm gal}\rangle(m_1) dm_1 \times\;\;\;\;\;\;\;\;\;\;\;\;\;\;\;\;\;\;
\;\;\;\;\;\;\;\;\;\;\\\int n(m_2) \langle N_{\rm gal} \rangle(m_2) dm_2
\int \rho_{m_1}({\bf x}-{\bf x}_1) d^3x_1\nonumber\\
\times\int \rho_{m_2}({\bf x}^\prime-{\bf x}_2) 
\xi({\bf x}_1-{\bf x}_2; m_1, m_2) d^3 x_2,\;\;\;\;\;\;\;\;\;\,\nonumber
\end{eqnarray}
where the first term $\xi_g^{1h}$ accounts for pairs of galaxies residing 
within 
the same halo, while the $\xi_g^{2h}$ represents the contribution coming 
from galaxies in different haloes. Note that all the above quantities are 
dependent on the redshift $z$, even though we have not made it explicit.\\
In the above equations, $\langle N_{\rm gal} \rangle (m)$ is the mean number 
of galaxies 
per halo of mass $m$, and $\langle N_{\rm gal} (N_{\rm gal}-1)\rangle (m)$ 
-- also dependent 
on the mass 
of the halo hosting the galaxies -- is a measure of the spread about this mean 
value. The mean comoving number density of galaxies is defined as:
\begin{eqnarray}
\bar{n}_g=\int n(m) \langle N_{\rm gal} \rangle(m) dm,
\label{eq:avern}
\end{eqnarray}
where $n(m)$ is the halo mass function which gives the number density 
of dark matter 
haloes per unit mass and volume. $\xi({ \bf x}_1-{\bf x}_2; m_1, m_2)$ is 
the two-point cross-correlation function between haloes of mass 
$m_1$ and $m_2$ and, 
finally, $\rho_m({\bf y})$ is the (spatial) density distribution of galaxies 
within the haloes, normalized so to obtain 
\begin{eqnarray}
\int_0^{r_{\rm cut}} \rho_m({\bf y})d^3y=1,
\label{eq:rho}  
\end{eqnarray}
where $r_{\rm cut}$ is the radius which identifies the outer boundaries of the 
halo.\\

\noindent
From the above discussion it then follows that, in order to 
work out $\xi_g$, we need to specify the halo-halo 
correlation function, the halo mass function, the spatial distribution 
of galaxies within the haloes and a 
functional form for the number distribution of galaxies within the haloes. This is 
done as follows.
\subsection{Halo-Halo Correlation Function}
An approximate model for 
the 2-point correlation function of dark--matter haloes can be easily 
obtained from the mass autocorrelation function as 
(see e.g. Porciani \& Giavalisco, 2002)
\begin{eqnarray}
\xi({\bf r},z,m_1,m_2)=\;\;\;\;\;\;\;\;\;\;\;\;\;\;\;\;\;\;\;\;\;\;\;\;\;\;\;
\;\;\;\;\;\;\;\;\;\;\;\;\;\;\;\;\;\;\;\;\;\;\;\;\;\;\;
\nonumber\\=\left\{\begin{array}{ll} 
\xi_{dm}(r,z)b_1(m_1,z)b_2(m_2,z) \;\;\;{\rm if}\; r\ge r_1+r_2\\
-1 \;\;\;\;\;\;\;\;\;\;\;\;\;\;\;\;\; {\rm otherwise}, 
\end{array}\right.
\label{eq:xihalo}
\end{eqnarray}
where the above expression takes into account the halo-halo spatial exclusion 
($r_1=r_{\rm cut_1}$ and $r_2=r_{\rm cut_2}$ are the Eulerian radii of the 
collapsed haloes, in general 
identified with their virial radii), and the mass-mass correlation function 
$\xi_{dm}(r,z)$ -- fully specified for a given cosmological model and a chosen 
normalization of $\sigma_8$ -- has been calculated  
following the approach of Peacock \& Dodds (1996) 
which is sufficiently accurate both in the linear and non-linear regimes.
In fact, results obtained with the more precise algorithm developed by 
Smith et al. (2003) significantly differ from those obtained with
the method by Peacock \& Dodds (1996) 
only in the regime where the 1-halo term dominates the 2-point clustering 
signal.

The linear bias factor $b(m,z)$ of individual haloes of mass $m$ at 
redshift $z$ can instead be written as (Sheth \& Tormen, 1999;
see also Cole \& Kaiser 1989; Mo \& White 1996; Catelan et al. 1997;
Porciani et al. 1998)
\begin{eqnarray}
b(m,z)=1+\frac{a\nu-1}{\delta_c}+\frac{2p/\delta_c}{1+(a\nu)^p},
\label{eq:biasst}
\end{eqnarray}  
with $p=0.3$, $a=0.707$, $\nu=(\delta_c/\sigma)^2$, where 
$\delta_c\simeq 1.686$ 
and $\sigma$ respectively are the critical overdensity for collapse  
and the linear rms variance of the power spectrum on the mass scale $m$ at 
redshift $z$. 

Note that, at variance with previous works (e.g. Peacock \& Smith, 2000; 
Seljak, 2000; Scoccimarro et al.,2001; Berlind \& Weinberg, 2001) where
the halo-halo correlation function was only derived in the linear regime,
our equation (\ref{eq:xihalo})
fully accounts for the non-linear evolution of density fluctuations.
Using linear theory to compute the 2-halo term
can be considered as a good approximation on large scales ($r\simgt 5$~Mpc) 
-- where the clustering growth 
is indeed still linear -- and does not create any problems 
on small scales ($r\simlt 1$~Mpc) -- where the 1-halo 
term $\xi_g^{1h}$ dominates the clustering signal. It, however, breaks down 
at intermediate distances where the $\xi_g^{1h}$ and $\xi_g^{2h}$ 
contributions 
are of comparable importance. It follows that the use of a linear halo-halo 
correlation function in equation (3) systematically leads to a 
serious underestimate of the clustering signal (1) produced by low-$z$ 
galaxies on scales 
$1\simlt r/[\rm Mpc]\simlt 5$, when compared with results obtained by  
taking into account the fully non-linear behaviour of $\xi_{dm}(r,z)$. 
For this reason, 
we believe our approach to be more consistent than the ones adopted so far. 
We note that similar models -- unknown to us till the very last stages of the 
present paper -- have also been used by Zehavi et al. (2003), 
Yang et al. (2003) and van den Bosch et al. (2003). 

\subsection{Mass Function}
For the analytical expression of the halo mass function we rely once again 
on the Sheth \& Tormen (1999) form:
\begin{eqnarray}
n(m,z)= \frac{A\bar{\rho}}{m^2} \sqrt{\frac{a\nu}{2\pi}} 
\left(1+\frac{1}{(a\nu)^p}\right) 
\exp{\left(-\frac{a\nu}{2}\right)}
\left|\frac{d({\rm ln}\nu)}{d({\rm ln} {\it m})}\right|
\label{eq:nm}
\end{eqnarray}
(with $A=0.322$, $\bar{\rho}$ the mean background density and the other 
quantities defined as above), since it gives an accurate fit to the 
results of N-body simulations with the same initial conditions
(Jenkins et al. 2001; Sheth, Mo \& Tormen 2001).

\subsection{Spatial Distribution of Galaxies}
The first, easiest approach one can take for the spatial distribution of 
galaxies within a halo of specified mass $m$ is to assume that galaxies 
follow the dark matter profile. Under this hypothesis we can then write
\begin{eqnarray}
\rho^\prime_m(r)=\rho_m(r)\cdot m=
\bar{\rho}\;\frac{f c^3/3}{cr/r_{\rm vir}(1+cr/r_{\rm vir})^2},
\label{eq:NFW}
\end{eqnarray}
where we use the Navarro, Frenk \& White (1997, hereafter NFW) expression, 
which provides a good description of the density distribution within 
virialized haloes in numerical simulations. In equation (\ref{eq:NFW}), 
$r_{\rm vir}$ is the virial radius of the halo, related to its mass via 
$m=(4\pi r_{\rm vir}^3/3)\Delta\bar{\rho}$, where $\Delta$ (=340 for an 
$\Omega=0.3$ universe at $z=0$) is the characteristic density contrast of 
virialized systems; 
$f=\Delta/[\rm{ln(1+c)-c/(1+c)}]$, and for the concentration parameter $c$ we 
use equations (9) and (13) in Bullock et al. (2001). 

Clearly, the assumption for the distribution of galaxies within 
virialized haloes to trace the dark matter profile is not necessarily true. 
For instance, the semi-analytic models by Diaferio et al. (1999) suggest that 
this cannot hold for both late-type and early-type galaxies since 
blue galaxies tend to reside in the outer regions of their parent haloes, 
while red galaxies are preferentially found near the halo centre.\\
For this reason, in the following analysis we will also consider spatial 
distributions of the form $\rho_m^\prime(r)\propto (r/r_{\rm vir})^{-\beta}$,
with $\beta=2,2.5,3$, where the first value corresponds to the singular 
isothermal sphere case. 

The last remark concerns the choice for values of $r_{\rm cut}$ in equations 
(\ref{eq:rho}) and (\ref{eq:xihalo}). All the profiles (both the NFW and 
the power-laws) considered so far formally extend to infinity, leading to 
divergent values for the associated masses. 
This implies the need to ``artificially'' truncate the distribution profiles 
at some radius $r_{\rm cut}$. One sensible choice is to set $r_{\rm cut}\equiv 
r_{\rm vir}$, since one expects galaxies to form within virialized regions, 
where the overdensity is greater than a certain threshold. However, this might 
not be the only possible choice since for instance -- as a consequence of 
halo-halo merging -- galaxies might also be found in the outer regions of the 
newly-formed halo, at a distance from the center greater than $r_{\rm vir}$.

\begin{figure}
\vspace{9 cm} \includegraphics{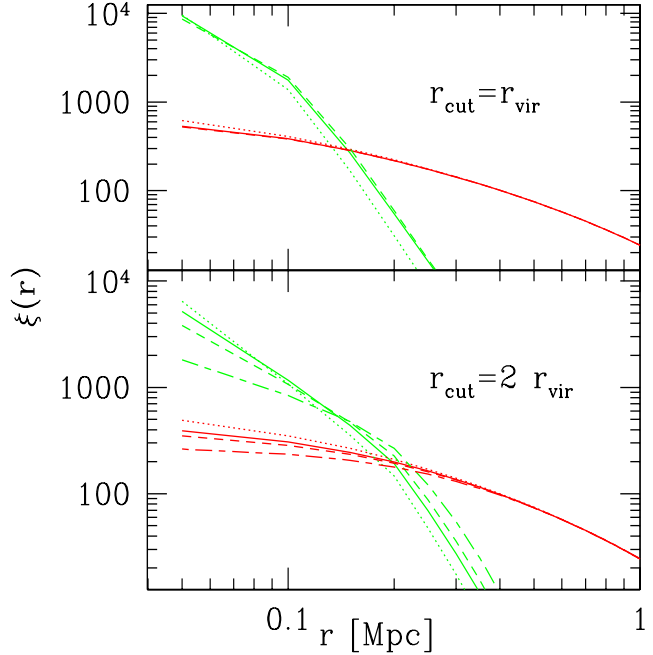}
\caption{Results for the galaxy-galaxy correlation function as obtained 
from different assumptions for the galaxy distribution profiles $\rho(r)$ 
and different choices for $r_{\rm cut}$. The case for $r_{\rm cut}=
r_{\rm vir}$ is shown in the top panel, while the one for 
$r_{\rm cut}=2\cdot r_{\rm vir}$ is presented in the bottom panel. Solid, 
short-dashed, long-short-dashed and dotted lines represent the predictions 
for a NFW, a power-law with $\beta=2.5$, a power-law with $\beta=2$ and 
power-law with $\beta=3$ distribution runs. Lower curves correspond 
to the term $\xi_g^{2h}$ in equation (\ref{eq:xi}) (contribution from objects 
in different haloes), while the upper ones indicate the $\xi_g^{1h}$ term. 
All the above curves have been calculated by setting in equation (\ref{eq:xi}) 
$\langle N_{\rm gal}\rangle=\langle N_{\rm gal}(N_{\rm gal}-1)\rangle=1$ 
for all 
halo masses greater than $10^{10.7} m_{\sun}$ and 0 otherwise.
\label{Fig:xi}}
\end{figure}

The way different assumptions for the steepness 
of the profiles and different choices for $r_{\rm cut}$ affect the 
galaxy-galaxy correlation function on small scales ($r \simlt 1$~Mpc) 
is presented in Figure 
(\ref{Fig:xi}). To this particular aim, both $\xi_g^{1h}$ and $\xi_g^{2h}$ 
have been derived 
from equation (\ref{eq:xi}) by setting $\langle N_{\rm gal}\rangle=
\langle N_{\rm gal} (N_{\rm gal}-1)\rangle=1$ 
for all halo masses greater than $10^{10.7} m_{\sun}$ and 0 otherwise.\\ 
The two panels show the case for $r_{\rm cut}=
r_{\rm vir}$ (top) and $r_{\rm cut}=2\cdot r_{\rm vir}$ (bottom), 
while solid, 
short-dashed, long-short-dashed and dotted lines respectively 
represent the results for 
a NFW, a power-law with $\beta=2.5$, a power-law with $\beta=2$ and 
a power-law with $\beta=3$ distribution profiles. Lower curves 
(for $r\to 0$) correspond 
to the $\xi_g^{2h}$ term (contribution from objects 
in different haloes), while the upper ones indicate the $\xi_g^{1h}$ term.\\ 
As Figure (\ref{Fig:xi}) clearly shows, both the scale at which the 
transition from a regime 
where objects in different haloes dominate the clustering signal  
to a regime where galaxies within the same halo start giving a contribution 
and the amplitudes of the $\xi_g^{1h}$ and $\xi_g^{2h}$ terms  
greatly depend on the radius chosen to truncate the distribution profiles. 
More in detail, the amplitude of both terms decreases and differences 
between predictions 
obtained for different profiles become more pronounced as the value 
for $r_{\rm cut}$ increases. Finally note that -- independently of the value 
of $r_{\rm cut}$ -- a stronger clustering signal at small scales is in general 
expected for steeper density profiles. 

Results originating from different assumptions for a spatial distribution of 
galaxies within dark matter haloes and different  
truncation radii will be further investigated in the following sections.

\subsection{Halo Occupation Function}
A key ingredient in the study of the clustering properties of galaxies
is their halo occupation function $p(N_{\rm gal}|m)$ which gives the 
probability 
for a halo of specified mass $m$ to contain $N_{\rm gal}$ galaxies. In the 
most 
general case, $p(N_{\rm gal}|m)$ is entirely specified by the knowledge of its 
$n$
moments $\langle N^n_{\rm gal}\rangle(m)$ which in principle can be observationally 
determined 
by means of the so-called ``counts in cells'' analysis (see e.g. Benson, 
2001). Unfortunately this is not feasible in reality, as measures of 
the higher moments of the galaxy distribution get extremely noisy 
for $n>4$ even for 2-dimensional catalogues (see e.g. Gaztanaga, 1995 
for an analysis of the APM survey).\\
A possible way to overcome this problem is to rely on the lower-order 
moments of the galaxy distribution to determine the low-order 
moments of the halo occupation function, and then assume a functional form 
for $p(N_{\rm gal}|m)$ in order to work out all the higher moments (see e.g. 
Scoccimarro et al., 2001; Berlind \& Weinberg, 2002). Clearly, 
better and better determinations of $p(N_{\rm gal}|m)$ are obtained as we 
manage 
to estimate higher and higher moments of the galaxy distribution function.

Since this work relies on measurements of the two-point correlation function 
of 2dF galaxies, equation (\ref{eq:xi}) shows that in our case only 
the first and second moment of the halo occupation function can be 
determined from the data, as these are the only two quantities which 
play a role in the theoretical description of $\xi_g$.\\
Following Scoccimarro et al. (2001), we chose to write the mean number of 
galaxies per halo of specified mass $m$ as
\begin{eqnarray}
\langle N_{\rm gal} \rangle(m) =0\;\;\;\;\;\;\;\;\;\;\;\;\;\;\;\;\;\;\;\;\;\;\;\;\;\;\;\;\;\;
\rm {if} \;\;{\it m<m}_{\rm cut}\nonumber \\
\langle N_{\rm gal} \rangle(m) =(m/m_0)^{\alpha_1}\;\;\;\;\;\;\rm{if}\;\; {\it m}_{\rm cut}\le
{\it m< m}_0 \\
\langle N_{\rm gal} \rangle(m)=(m/m_0)^{\alpha_2}\;\;\;\;\;\;\;\;\;\;\;\;\;\;\;
\rm{if} \;\;{\it m\ge m}_0, \nonumber
\label{eq:Ngal},
\end{eqnarray}
where $m_{\rm cut}$, $m_0$, $\alpha_1$ and $\alpha_2$ are parameters to be 
determined by comparison with observations. The choice of the above functional 
form for $\langle N_{\rm gal}\rangle(m)$ relies on the physics connected with galaxy 
formation processes (see e.g. Benson et al., 2001; Somerville et al., 2001; 
Sheth \& Diaferio, 2001). For instance, $m_{\rm cut}$ gives the minimum mass 
of a halo able to host a galaxy since -- for potential wells which are not 
deep enough -- galaxy formation 
is inhibited by supernova processes occurring amongst the first stars 
which can blow the remaining gas away from the halo itself 
therefore suppressing further star-formation. On the other hand, given that 
the internal velocity dispersion of haloes increases with halo mass, 
gas is expected to cool less efficiently and therefore inhibit at some level 
galaxy formation in more massive haloes. In a hierarchical scenario for
structure formation, however, the most massive objects are formed by merging
and accretion of smaller units. Thus, one expects the number of galaxies
to increase with the halo size in the high-mass regime.
All this can then be parameterized by a broken 
power-law of the form (10), 
with $m_0$ the ``threshold mass'' at which the transition between the
two different scaling laws occurs. 

The last ingredient needed for 
the description of 2-point galaxy clustering is the second moment of the halo
occupation function appearing in equation (\ref{eq:xi}).
This term quantifies the
spread (or variance) about the mean value of the number counts 
of galaxies in a halo. 
A convenient parameterization for this quantity is:
\begin{eqnarray}
\label{second}
\langle N_{\rm gal}(N_{\rm gal}-1)\rangle(m)=\alpha(m)^2 \langle N_{\rm gal}\rangle^2,
\end{eqnarray} 
where $\alpha(m)=0,{\rm log}(m/m_{\rm cut})/{\rm log}(m_0/m_{\rm cut}),1$, 
respectively 
for $m< m_{\rm cut}$, $m_{\rm cut}\le m < m_0$ and $m\ge m_0$. Note that, 
while the high-mass value for $\alpha(m)$ simply reflects a Poissonian 
statistics, 
the functional form at intermediate masses (chosen to fit the results from 
semi-analytical models -- see e.g. Sheth \& Diaferio, 2001; 
Berlind \& Weinberg 2002 --
and smoothed particle hydrodynamics simulations -- Berlind et al. 2003) 
describes the sub-Poissonian regime.\\
\\
\\
\noindent
As a last consideration, 
note that equation (\ref{eq:xi}) presents in both the 
$\xi_g^{1h}$ and $\xi_g^{2h}$ terms convolutions of density 
profiles. Since this is somehow difficult to deal with, we prefer to 
work in Fourier space where all the expressions simply become 
multiplications over the Fourier transforms of the profiles. 
Equation (\ref{eq:xi}) is therefore equivalent to:
\begin{eqnarray}
\bar{n}^2_g \Delta_g(k)=\int n(m) 
\langle N_{\rm gal}(N_{\rm gal}-1)\rangle(m) |u_m(k)|^2 dm + \nonumber\\
\int u_{m_1}(k) \langle N_{\rm gal}\rangle(m_1) n(m_1) dm_1\times\;\;\;\;\;\;\;\;\;\;\;\;\;
\;\;\;\;\nonumber\\
\int u_{m_2}(k) n(m_2) \langle N_{\rm gal}\rangle(m_2) \Delta(k,m_1,m_2) dm_2 ,
\label{eq:pk}
\end{eqnarray}
where -- allowing for the exclusion effects as in equation (6) -- 
$\Delta(k,m_1,m_2)=\Delta_{dm}(k) b(m_1) b(m_2)$ is the power spectrum 
of haloes of mass $m_1$ and $m_2$ ($\Delta_{dm}(k)=k^3/(2\pi^2) P_{dm}(k)$ 
is the 
normalized non-linear
power-spectrum for dark matter -- (see e.g. Peacock \& Dodds, 1996),
\begin{eqnarray}
u_m(k)=\frac{k^3}{(2\pi^2)}\int_0^{r_{\rm cut}}\rho_m(r) {\rm sin}(kr)/(kr) 
4\pi r^2 dr,
\end{eqnarray}
with $\rho_m(r)$ defined as in (\ref{eq:rho}), is the Fourier transform of 
the galaxy distribution profile truncated at $r_{\rm cut}$, and where all the 
quantities are implicitly taken at a fixed $z$. In this framework, the 
galaxy-galaxy correlation 
function can then be obtained from equation (\ref{eq:pk}) via
\begin{eqnarray}
\xi_g(r)=\int \Delta_g(k)\frac{{\rm sin}(kr)}{kr} \frac{dk}{k}.
\end{eqnarray}  

\section{The 2dF Galaxy Redshift Survey}

The 2dF Galaxy Redshift Survey (2dFGRS: Colless et al., 2001) 
is a large-scale survey aimed at obtaining spectra for 250,000 galaxies 
to an extinction-corrected limit for completeness of $b_J=19.45$ over 
an area of 2151 square degrees. 
The survey geometry consists of two broad declination strips, a larger one 
in the SGP covering the area $3^h 30^m\simlt {\rm RA}({\rm 2000})\simlt 
21^h40^m$, $-37.5^\circ \simlt {\rm dec}({\rm 2000})\simlt -22.5^\circ$ and a 
smaller one 
set in the NGP with $9^h 50^m\simlt {\rm RA}({\rm 2000})\simlt 14^h50^m$, 
$2.5^\circ \simlt{\rm dec}({\rm 2000}) \simlt -7.5^\circ$, plus 100 random 
2-degree 
fields spread uniformly over the 7000 square degrees of the APM catalogue in 
the southern Galactic hemisphere.

The input catalogue for the survey is a revised version of the APM galaxy 
catalogue (Maddox et al. 1990a, 1990b, 1996) which includes over 5 
million galaxies down to $b_J=20.5$. 
%
Redshifts for all the sources brighter than $b_J=19.45$ are determined
in two independent ways, via both cross-correlation of the spectra
with specified absorption-line templates (Colless et al., 2001) and by
emission-line fitting. These automatic redshift estimates have then
been confirmed by visual inspection of each spectrum, and the more
reliable of the two results chosen as the final redshift. A quality
flag was assigned to each redshift: $Q=3$, $Q=4$ and $Q=5$ correspond to 
reliable redshift determination, $Q=2$ means a probable redshift and $Q=1$
indicates no redshift measurement.  The success rate in redshift
acquisition for the surveyed galaxies (determined by the inclusion in
the 2dF sample of only those objects with quality flags $Q=3$ to $Q=5$)
is estimated about 95 per cent (Folkes et al., 1999).  The median
redshift of the galaxies is 0.11 and the great majority of them have
$z<0.3$. 

\subsection{Luminosity Functions and Number Densities}
Madgwick et al. (2002) calculate the optical $b_j$ luminosity function (LF) 
for different subsets of $\rm{M}-5\rm{log_{10}}(h_0)\le-13$ 2dFGRS 
galaxies defined by their spectral type.
The spectral classification -- based upon a Principal Component Analysis -- 
was performed for 75,589 galaxies found at redshifts $0.01 < z < 0.15$ and  
allowed to divide the whole population into four well-defined classes 
according to the strength of their star-formation activity:
from type 1 (early-type galaxies only showing absorption lines in their 
spectra) to type 4 (extremely active star-forming galaxies).
 
Fitting functions for the different luminosity distributions are presented in 
Madgwick et al. (2002) and can be used to determine the average 
number density of galaxies of different spectral types via 
(see e.g. Lin et al., 1996):
\begin{eqnarray}
\bar{n}_{g_i}=N_i\times 
\left[\int_{z_{\rm min}}^{z_{\rm max}}S_i(z)(dV/dz)dz\right]^{-1},
\label{eq:ng}
\end{eqnarray}
with the selection function $S_i(z)$ defined as
\begin{eqnarray}
S_i(z)=\frac{\int^{\rm{min[M^{\it i}_{max}(z)-M_{max}]}}
_{\rm{max[M^{\it i}_{min}(z)-M_{min}]}}\Phi_i(\rm{M}) d{\rm M}}
{\int_{\rm{M_{min}}}^{\rm{M_{max}}} \Phi_i(\rm{M}) d{\rm M}},
\label{eq:sz}
\end{eqnarray}
where $dV/dz$ is the volume element, $i=1,2,3,4$ refers to the spectral 
class, $N_i$ is the 
total number of galaxies belonging to a specific type found in the survey and  
$\Phi_i$ is their luminosity function. $\rm {M_{min}=-13+5 log_{10}}(h_0)$ , 
$z_{\rm min}=0.01$ and $\rm {M_{max}\simeq -21+5 log_{10}}(h_0)$, 
$z_{\rm max}=0.15$ 
respectively are the minimum and maximum absolute magnitudes and redshifts 
of the objects under exam, 
while ${\rm M^{\it i}_{min(max)}}(z)=b_{\rm max(min)}-25-5 {\rm log} 
(d_L^i)-k_i(z)$, where $k_i(z)$ is the K-correction, $d_L^i$ 
the luminosity distance and $b_{\rm min}=14$, $b_{\rm max}=19.45$ are 
the apparent magnitude limits of the 2dF survey.
 
We have then applied equation (\ref{eq:ng}) to the four different galaxy types 
and found:
\begin{eqnarray}
\bar{n}_{g_1}= 9.6 \cdot 10^{-3}\left\{\begin{array}{ll}
-4.4\cdot10^{-3}\\
+7.0\cdot10^{-2} 
\end{array} 
\right .\nonumber \\
\bar{n}_{g_2}=0.0103\pm 3\cdot 10^{-4} \;\;\;\;\;\;\;\;\;\;\;\;\;\;
\nonumber\\ 
\bar{n}_{g_3}=0.015 \left\{\begin{array}{ll}
-8\cdot10^{-4}\\
+9\cdot10^{-4} 
\end{array} 
\right .\;\;\;\;\;\;\;\;\; \\
\bar{n}_{g_4}=0.018 \pm 1.8\cdot10^{-3},\;\;\;\;\;\;\;\;\;\;\;\;\nonumber
\label{eq:ngobs}
\end{eqnarray}
where the quoted 1$\sigma$ errors, in all but the type 1 case, are obtained by 
varying the 'break' luminosities ${\rm M}^*_i$
and the faint-end slopes $\mu_i$ of the Schechter functions $\Phi_i(\rm M)=
(0.4\:{\rm ln}10)\Phi^*_i(10^{-0.4({\rm M-M^*_{\it i}})})^{1+\mu_i}\rm{exp}
(-10^{-0.4(M-M^*_{\it i})})$  
(note that our calculations are independent of the normalizations $\Phi^*_i$) 
along their joint 1$\sigma$ error ellipse.

Two features have to be noticed in (17). First, errors on 
$\bar{n}_{g_1}$ -- especially the upper 1$\sigma$ limit of the number 
density -- are significantly larger than those derived for the other three 
classes of sources. This is due to the fact that the Schechter function 
might not provide a good fit to the faint end of the LF of 
type 1 galaxies (Madgwick et al., 2002); an extra term of the form 
$\Phi({\rm M})=10^{a+b\rm{M}}$ is needed for 
$\rm{M-5log_{10}}(h_0)\simgt -16$, where 
the parameters $a$ and $b$ can only be determined from the data with quite 
large uncertainties.
Since it is the faint end of the LF which mostly contributes to the 
determination of $\bar{n}_{g_1}$, and this is the region where the errors on 
$a$ and $b$ dominate over those derived for the various parameters in the 
Schechter function, this explains our finding for such large uncertainties 
associated to the measurement of $\bar{n}_{g_1}$.\\
The second point to be noticed concerns the errors associated to 
$\bar{n}_{g_4}$. In fact it turns out that, also in the case of type 4 
galaxies, the Schechter function does not provide a good fit to the faint 
end of the luminosity function as it systematically overestimates the number 
density of $\rm{M-log_{10}}(h_0)\simgt -16$ galaxies. This implies that 
the total number of type 4 sources as derived from integration of the LF is 
in agreement with the observed one only 
if we subtract to the best estimate $\bar{n}_{g_4}$ in eq. 
(17) an error corresponding to a 3$\sigma$ confidence level in 
$\mu_4$. Since, as we will see better in the next Section, 
the observed number density of galaxies plays a relevant role in the 
determination of the best halo occupation model, we have then decided in the 
case of type 4 sources to consider errors on $\mu_4$ at the 3$\sigma$ 
level; this propagates to a lower limit for $\bar{n}_{g_4}=0.0127$. 

Finally, from the luminosity function we can also determine 
another quantity which will be useful in the following Sections: the effective 
redshift of a class of sources defined as 
$z_i^{\rm eff}=\int_{z_{\rm min}}^{z_{\rm max}}
z S_i(z) (dV/dz) dz$. Numbers obtained for the different types of galaxies 
under exam then read: 
$z_1^{\rm eff}=0.098$, $z_2^{ \rm eff}=0.091$, $z_3^{\rm eff}=0.082$, 
$z_4^{\rm eff}=0.078$, 
indicating a preference for late-type galaxies to be found at lower 
redshifts than early-type objects.

\subsection{Correlation Functions} 
A sample with the same selection criteria as the one introduced in Section 3.1 
but containing more (96,791) objects has been used by Madgwick et al. 
(2003) to calculate the clustering properties of galaxies belonging to 
different spectral types. In order to increase the statistics 
associated to the measurements, galaxies have been grouped into two broad 
categories: early-types -- 36318 objects with effective redshift 0.1 to be 
identified with those sources belonging to spectral class 1 -- and late-type 
-- 60473 objects with effective redshift 0.09 obtained by taking into account 
all galaxies from spectral classes 2 3 and 4 -- ones.    

The different observed correlation functions are shown in Figures 
(\ref{Fig:xiearly}) and (\ref{Fig:xilate}) for early-type 
and late-type galaxies, respectively.
In order to get rid of redshift distortions, the correlation function in real 
space has been inferred by computing the bi-dimensional correlation parallel
and transverse to the line-of-sight, $\xi_g(r_P,r_T)$, and by integrating it
in the $r_P$ direction. The quantity presented in 
Figures (\ref{Fig:xiearly}) and (\ref{Fig:xilate}) therefore corresponds to
\begin{eqnarray}
\bar{\xi}_g(r_T)=2\int_{r_T}^{\infty}\xi_g(r)\frac{r dr}{(r^2-r_T^2)^{1/2}}.
\label{eq:xibar}
\end{eqnarray}      
Error-bars have been obtained by bootstrap resampling, adapting the
method presented by Porciani \& Giavalisco (2002).\\

Measurements for the integrated correlation function 
(\ref{eq:xibar}) differ for the two populations not only in their amplitude,
but also in the slopes (Madgwick et al., 2003 -- by fitting the data 
with a power-law $\xi(r)=(r/r_0)^{\gamma}$ -- find $r_0=3.67\pm 0.30$
~h$^{-1}$~Mpc, $\gamma=1.60\pm 0.04$ for late-type galaxies and 
$r_0=6.10\pm 0.34$~h$^{-1}$~Mpc, $\gamma=1.95\pm 0.03$ for early-type 
sources). As we will 
extensively see in the next Sections, these differences can provide a great 
amount of information on the processes associated to galaxy formation  
in the two different cases of passive and active star-forming galaxies.

\section{Results}
In order to determine the best values for the parameters describing the halo 
occupation number (10) we allowed them to vary within the following region:
\begin{eqnarray}
-1 \le \alpha_1\le 2;\;\;\;\;\;\;\;\;\;\;\;\;\;\;\nonumber\\
 -1\le \alpha_2\le 2;\;\;\;\;\;\;\;\;\;\;\;\;\;\;\nonumber\\ 
10^9\; m_\odot\le m_{{\rm cut}} \le 10^{13}\; m_\odot; \nonumber \\ 
m_{{\rm cut}}\le m_0\le m_{{\rm cut}}\cdot 10^3 .\;\;\;\nonumber 
\end{eqnarray}
Combinations of these four quantities have then been used to evaluate 
the mean number density of galaxies $\bar{n}_g$ via equation ({\ref{eq:avern}).
Only values for $\bar{n}_g$ within 2$\sigma$ from the observed ones 
(quoted in Section 3.1) were accepted and the corresponding values for 
$\alpha_1$, $\alpha_2$, $m_{\rm cut}$ and $m_0$ have subsequently been 
plugged into equation (\ref{eq:xi}) to produce -- for a specified choice of 
the distribution profile and effective redshift $z^{\rm eff}$ (slightly 
different in the case of early-type and late-type galaxies) -- the predicted 
galaxy-galaxy correlation function to be integrated via eq. (\ref{eq:xibar}) 
and compared with the data by means of a least squares ($\chi^2$) fit.\\

The value for the truncation radius of the halo 
was set to $r_{\rm vir}$ and the above procedure repeated for different 
choices of $\rho_m(r)$. The following sub-sections describe the results 
obtained for the two different classes of Early- and Late-type galaxies.

\subsection{Early-type galaxies}
The best description of the data in this case is provided by a model with  
$\alpha_1\simeq -0.2$, $\alpha_2=1.1$, $m_{\rm cut}=10^{12.6}m_\odot$,
$m_0=10^{13.5}m_\odot$ (in good agreement with the findings of Zehavi et al., 2003) 
and with a mild preference for galaxies to be distributed 
within their dark matter haloes according to a NFW profile. 
The projected galaxy-galaxy correlation function for this combination of 
values and the NFW spatial distribution is illustrated by the solid curve in Figure 
(\ref{Fig:xiearly}), while the dashed and dotted lines respectively indicate 
the contribution $\xi_g^{2h}$ from galaxies residing in different haloes and 
the $\xi_g^{1h}$ term originating from galaxies within the same halo. The 
agreement between data and predictions is good at all scales even though 
the model tends to underestimates the correlation function at
intermediate scales (between 3 and 10 Mpc), where the maximum discrepancy is 
$\sim 20$ per cent.
\\

\begin{figure}
\vspace{8 cm} \includegraphics{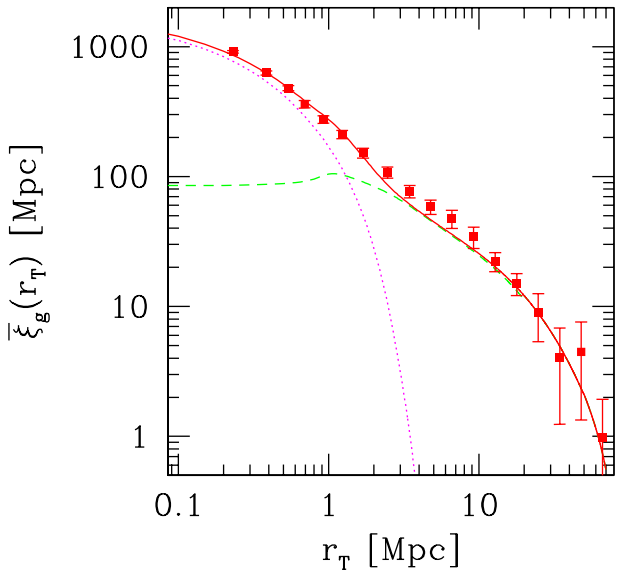}
\caption{Projected correlation function of early-type galaxies. Data-points 
represent the results from Madgwick et al. (2003), while the solid curve 
is the best fit to the measurements obtained for a halo number density of the 
form (10), with $\alpha_1=-0.2$, $\alpha_2=1.1$, $m_{\rm cut}=10^{12.6}m_\odot$,
$m_0=10^{13.5}m_\odot$ and for galaxies distributed within their dark matter 
haloes according to a NFW profile. Dashed 
and dotted lines respectively indicate 
the contribution $\xi_g^{2h}$ from galaxies residing in different haloes and 
the $\xi_g^{1h}$ term originating from galaxies within the same halo.
\label{Fig:xiearly}}
\end{figure}

A more quantitative assessment of the goodness of the match can be found 
in Table~1  
which -- for each choice of the distribution profile $\rho_m(r)$ and for the 
different classes of early-type and late-type galaxies -- 
provides the value of the $\chi^2=\chi^2_{min}$ obtained for the best fit to 
the data and 
the corresponding estimates for the parameters which appear in the description 
of the halo occupation number (10). 1$\sigma$ errors on these quantities 
are obtained by requiring their different combinations not to produce models 
for the galaxy-galaxy correlation function which -- when compared to the 
measurements -- correspond to $\chi^2$ values which differ from the minimum 
by a factor greater than $\Delta\chi^2=3.53$, where this last figure has been 
derived for an analysis with three degrees of freedom
(assuming Gaussian errors). 
Three is in fact the 
number of degrees obtained if one subtracts to the number of independent  
$\bar{\xi}_g$ measurements (eight -- Darren Madgwick, private communication) 
\footnote{
It is in some sense arbitrary to decide how many principal components of
the correlated errors correspond to a real signal and how many correspond to 
noise. This can be done, for instance, by considering a fixed fraction of
the variance (see, e.g., Porciani \& Giavalisco 2002).}
the number of parameters to determine (four) and the (one) constraint on 
$\bar{n}_g$. 

\begin{figure}
\vspace{8 cm} \includegraphics{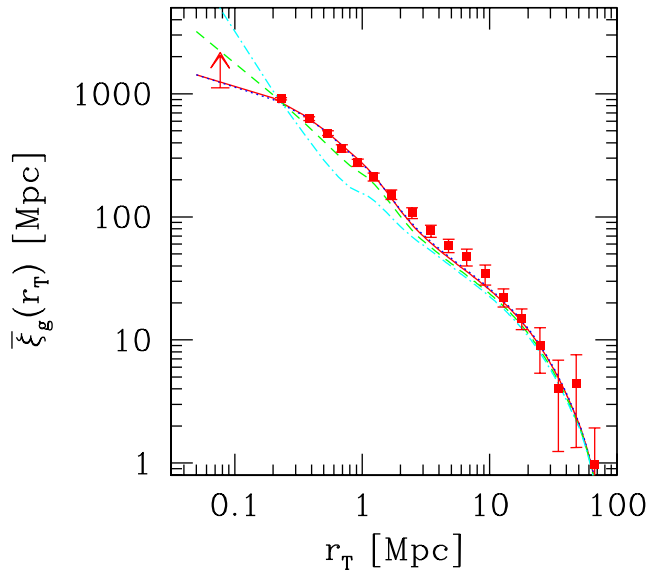}
\caption{Projected correlation function of early-type galaxies. Data-points 
represent the results from Madgwick et al. (2003), while solid, dashed, 
dotted and dashed-dotted curves illustrate the best-fit models respectively  
obtained for a NFW, a power-law with $\beta=2.5$, a power-law with $\beta=2$ 
and a power-law with $\beta=3$ galaxy distribution profiles 
(see text for details). The arrow shows the lower limit to the measured correlation 
function set on small scales by fibre collisions in the 2dF survey (Ed Hawkins, 
private communication).
\label{Fig:xiearlyall}}
\end{figure}

A closer look to Table~1 shows that, in the case of early-type galaxies, 
the best-fit values for the parameters appearing in equation (10) are 
independent -- except for the most extreme cases of very poor fits -- of the particular 
choice for the spatial distribution of galaxies 
within the haloes. This is verified even though different profiles are 
associated to different values for $\chi^2_{min}$ and is 
due to the fact that, while quantities such as $\langle N_{\rm gal}\rangle$ and 
$\langle N_{\rm gal}^2\rangle$ (and therefore the parameters associated to them) 
mainly determine the amplitudes of the $\xi_g^{1h}$ 
and $\xi_g^{2h}$ terms, the distribution of galaxies within the 
haloes is directly responsible for the slope of $\xi_g$, 
especially on scales $r\simlt 1$~Mpc.\\
This effect is better seen in Figure 
(\ref{Fig:xiearlyall}) which presents the best-fit models obtained for 
different choices of the distribution run. More in detail, the solid line 
is for a NFW profile, the dotted line for a power-law profile with 
$\beta=2$, the dashed line for a power-law profile with $\beta=2.5$ and the 
dashed-dotted line represents the case of a power-law profile with $\beta=3$.
Even though all the curves  
are almost indistinguishable from each other on all scales $r\simgt 2$~Mpc due to 
the very similar best-fit values obtained for the parameters which describe 
$\langle N_{{\rm gal}}\rangle$ and $\langle N_{{\rm gal}}(N_{{\rm gal}}-1)
\rangle$, this does not hold 
any longer at smaller distances. In fact, when we push the 
analysis inside the haloes, the profile assumes a 
crucial importance; for instance, in the case of early-type galaxies it is 
clear that, while profiles such as NFW and singular isothermal sphere 
can provide a good fit to the observations (with a slight preference for the first  
model), anything steeper than these two is drastically ruled out by the data since it 
does not exhibit enough power on scales $0.3\simlt r/[\rm{Mpc}]\simlt 2$.

Table~1 states the same conclusion from a more quantitative point of 
view showing that, while the NFW profile provides the best description of the data 
and the value of $\chi^2_{min}$ for a $\beta=2$ model is still 
within the 1$\sigma$ range of the acceptable fits (for Gaussian errors), 
profiles of the form  
power-law with $\beta\simgt 2.5$ are too steep to be accepted as satisfactory 
descriptions of the observed galaxy correlation function. We remark once again 
that, since the spatial distribution of galaxies within haloes of specified 
mass and their mean number and variance affect different regions of the 
observed correlation 
function, these two effects are untangled so that there is no 
degeneracy in the determination of the $\langle N_{\rm gal}\rangle$ and $\rho(r)$ 
preferred by the data.

Since different profiles lead to notably different predictions for the correlation 
function on increasingly smaller ($r\simlt 0.2$~Mpc, see Figure 3) scales, one would in 
principle like to be able to explore this region in order to put stronger constraints on 
the distribution of galaxies within their haloes. Unfortunately, this cannot be done 
because -- as Hawkins et al. (2003) have shown -- fibre collisions in the 2dF survey can 
significantly decrease the measured $\xi_g(r)$ on scales $r\simlt 0.1$~Mpc, leading 
to systematic underestimates.
Nevertheless, one can still use the 2dF measurements of the correlation 
function as a lower limit to the real galaxy-galaxy clustering strength, to be compared 
with predictions steaming from different density runs. This is done in Figure 3, where 
the 
arrow represents the 2dF measurement of the early-type correlation function on a scale 
$r\simeq 0.07$~Mpc. In this particular case, all the models appear to have enough power 
on such small scales as none of them falls below the accepted range of variability of 
the measured $\xi_g(r)$. Therefore, on the basis of the present data, we cannot break the 
degeneracy between NFW and PL2 profiles as to which one can provide the 
best description of the data in the whole $r$ range.

\begin{table*}
\begin{center}
\caption{Best-fit values for the parameters describing the halo occupation
number (10), expressed for different choices of the distribution
profile and truncation radius. PL2, PL2.5 and PL3 respectively correspond to 
profiles described by power-laws with slopes 
$\beta=2$, $\beta=2.5$ and $\beta=3$. Errors are for a 68.3\% confidence level 
and three degrees of freedom. $\bar{n}_g$ (in Mpc$^{-3}$) and b are the 
average number density 
of sources and the bias on large scales as derived from the combination of 
the best-fit parameters associated to each model. All the masses are measured in 
$m_{\odot}$ units.
Vanishing errorbars mean that either the best-fitting value lies on the
boundary of the sampled parameter space (e.g. when $\alpha_1=-1$)
or that the uncertainty is smaller than our grid step (0.1 for all
parameters).}

\begin{tabular}{lllllllllll}
\hline
\hline
EARLY-TYPE    & NFW & PL2 & PL2.5 & PL3 \\
\hline
$r_{\rm cut}=r_{\rm vir}$   &  $\chi_{min}^2=4.0$&$\chi_{min}^2=4.9$
     &$\chi_{min}^2=13.2$&$\chi_{min}^2=68.9$\\
  &  $\alpha_1=-0.2^{+0.6}_{-0.5}$&$\alpha_1=-0.5^{+1.0}_{-0.3}$&
    $\alpha_1=-0.6^{+0.7}_{-0.4}$&$\alpha_1=-0.5^{+0.1}_{-0.5}$\\
   & $\alpha_2=1.1^{+0.1}_{-0.2}$&$\alpha_2=1.0^{+0.1}_{-0.1}$&
    $\alpha_2=1.5^{+0.0}_{-0.2}$& $\alpha_2=1.8^{+0.0}_{-0.6}$\\
 &${\rm Log}[m_{\rm cut}]=12.6_{-0.5}^{+0.1}$&${\rm Log}[m_{\rm cut}]=
12.7_{-0.7}^{+0.0}$&${\rm Log}[m_{\rm cut}]=12.7_{-0.4}^{+0.2}$& ${\rm Log}
[m_{\rm cut}]=12.7^{+0.2}_{-0.2}$\\
  & ${\rm Log}[m_{0}]=13.5_{-0.5}^{+0.0}$&${\rm Log}[m_{0}]=13.4_{-0.4}
^{+0.1}$&
   ${\rm Log}[m_{0}]=13.7_{-0.3}^{+0.1}$&${\rm Log}[m_{0}]=13.9_{-0.4}^{+0.1}$\\
&$\rm n_g=8.4\cdot 10^{-4}$&$\rm n_g=8.6\cdot 10^{-4}$&
$\rm n_g=1.31\cdot 10^{-3}$&$\rm n_g=1.34\cdot 10^{-3}$\\
&b=1.33&b=1.34&b=1.31&b=1.30\\
\hline
\hline
LATE-TYPE    & NFW & PL2 & PL2.5 & PL3 \\
\hline
$r_{\rm cut}=r_{\rm vir}$ &  $\chi_{min}^2=90.4$&$\chi_{min}^2=107.6$
& $\chi_{min}^2=81.7$&$\chi_{min}^2=38.8$ \\
\hline
$r_{\rm cut}=2\;r_{\rm vir}$ &  $\chi_{min}^2=9.7$&$\chi_{min}^2=7.2$
& $\chi_{min}^2=10.7$&$\chi_{min}^2=14.4$&\\
      &$\alpha_1=-0.4^{+2.4}_{-0.6}$&$\alpha_1=-1.0^{+3.0}_{-0.0}$
      &$\alpha_1=-0.6^{+2.4}_{-0.4}$&$\alpha_1=-0.7^{+2.7}_{-0.3}$\\
      &$\alpha_2=0.7^{+0.1}_{-0.1}$&$\alpha_2=0.7^{+0.2}_{-0.1}$
      & $\alpha_2=0.7^{+0.3}_{-0.1}$&$\alpha_2=0.6^{+0.3}_{-0.1}$\\
      & ${\rm Log}[m_{\rm cut}]=11.0_{-1.8}^{+0.2}$& ${\rm Log}[m_{\rm cut}]
   =11.1_{-1.9}^{+0.1}$&${\rm Log}[m_{\rm cut}]=11.0_{-1.8}^{+0.3}$&
      ${\rm Log}[m_{\rm cut}]=11.0_{-1.8}^{+0.3}$\\
      & ${\rm Log}[m_{0}]=11.4_{-0.4}^{+0.6}$&${\rm Log}[m_{0}]=11.4_{-0.3}
      ^{+0.5}$ &${\rm Log}[m_{0}]=11.4^{+0.8}_{-0.4}$&
      ${\rm Log}[m_{0}]=11.4_{-0.4}^{+0.8}$\\
      &$\rm n_g=0.032$&$\rm n_g=0.031$ &$\rm n_g=0.034$&$\rm n_g=0.035$\\
      &b=0.98&b=0.99 & b=0.97&b=0.97\\
\hline
\hline

\end{tabular}
\end{center}
\end{table*}

If we then concentrate our attention on the two distribution profiles that can 
correctly reproduce the observations (NFW and power-law with $\beta=2$, 
hereafter PL2) and analyze the best-fit values obtained for the parameters 
describing the halo occupation number and their associated errors, we find 
for instance that, while the slope $\alpha_2$ -- which determines the 
increment of the number of sources hosted by dark matter haloes of increasing  
mass in the high-mass regime -- is very well determined, the situation is
more uncertain for what concerns $\alpha_1$, 
counterpart of $\alpha_2$ in the low-mass regime. On the other hand,
$m_{\rm cut}$ and $m_0$ exhibit errors of similar magnitudes, with upper 
limits better determined than the lower ones, this last effect possibly due 
to the constraints on $\bar{n}_g$ which discard every model not able to 
produce enough galaxies as it is the case for high values of $m_{\rm cut}$ and 
$m_0$.\\
An analysis of the $\chi^2$ hypersurface also shows that all the 
parameters but $\alpha_2$ (and especially $m_{\rm cut}$ and $m_0$) are 
covariant. This means that, in order to be consistent with the available data,
decreasing $m_{\rm cut}$ with respect to its best-fitting value
implies lowering $m_0$ and increasing $\alpha_1$.
The interplay is probably due to the fact that both $\alpha_1$ and 
$m_{\rm cut}$ only play a role in the low-mass regime which mainly affects 
the intermediate-to-large-scale regions of the galaxy correlation function, 
where measurements are more dominated by uncertainties. Conversely, $\xi_g$ on 
small scales is strongly dependent on the adopted value of $\alpha_2$, 
therefore making 
this last quantity measurable with an extremely high degree of precision. 

\subsection{Late-type galaxies}

\begin{figure}
\vspace{8 cm} \includegraphics{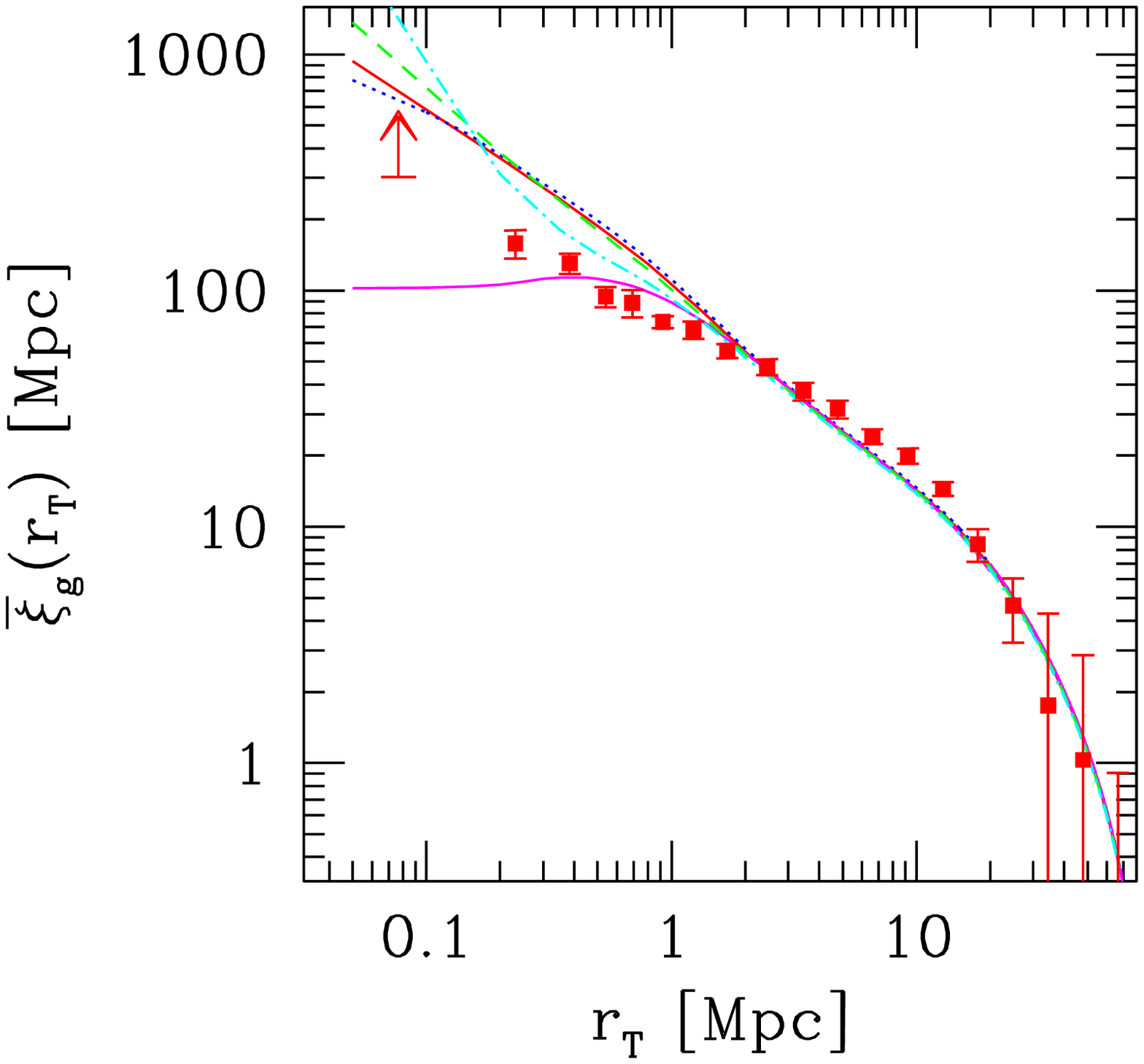}
\caption{Projected correlation function of late-type galaxies. Data-points 
represent the results from Madgwick et al. (2003), while solid, dashed, 
dotted and dashed-dotted curves illustrate the best-fit models respectively  
obtained for a NFW, a power-law with $\beta=2.5$, a power-law with $\beta=2$ 
and a power-law with $\beta=3$ galaxy distribution profiles and for a 
truncation radius $r_{\rm cut}=r_{\rm vir}$. The curve
flattening around $r\simlt 0.5$~Mpc 
illustrates the contribution $\xi_g^{2h}$ from galaxies residing in different 
haloes (see text for details). The arrow shows the lower limit to the measured 
correlation function set on small scales by fibre collisions in the 2dF survey 
(Ed Hawkins, private communication).
\label{Fig:xilateall}}
\end{figure}

\noindent
The case for late-type galaxies appears more tricky to treat in the framework 
of our analysis due to the shallow slope of the observed correlation function 
(see Section 3.2). In fact, as it can be appreciated in Figure 
(\ref{Fig:xilateall}),
no model can correctly describe the slow rise of the data on scales 
$r\simlt 2$~Mpc, even though the large-scale normalization of all the curves 
reproduces with a good approximation the measured one. This 
is shown in a more quantitative way in Table~1, which quotes the minimum 
$\chi^2$ values obtained for different distribution profiles and a truncation 
radius $r_{\rm cut}=r_{\rm vir}$: no model gives an acceptable fit and 
the corresponding figures for $\chi^2_{min}$ are -- in the best case -- 
around 39.\\
A closer look at Figure (\ref{Fig:xilateall}) indicates that the 
problem has to be connected with the excess of power on intermediate scales 
exhibited by the $\xi_g^{2h}$ contribution of pairs of galaxies from 
different haloes 
(solid curve which flattens on scales $\sim 0.5$~Mpc), which creates the 
mismatch between the observed slope and the steeper ``best-fit models''. In 
other words, what the plot reveals is that in the model there are too many 
pairs of galaxies coming from different haloes at distances 
$0.5\simlt r/[\rm Mpc]\simlt 2$ with respect to what the data seem to 
indicate. 

\begin{figure}
\vspace{8 cm} \includegraphics{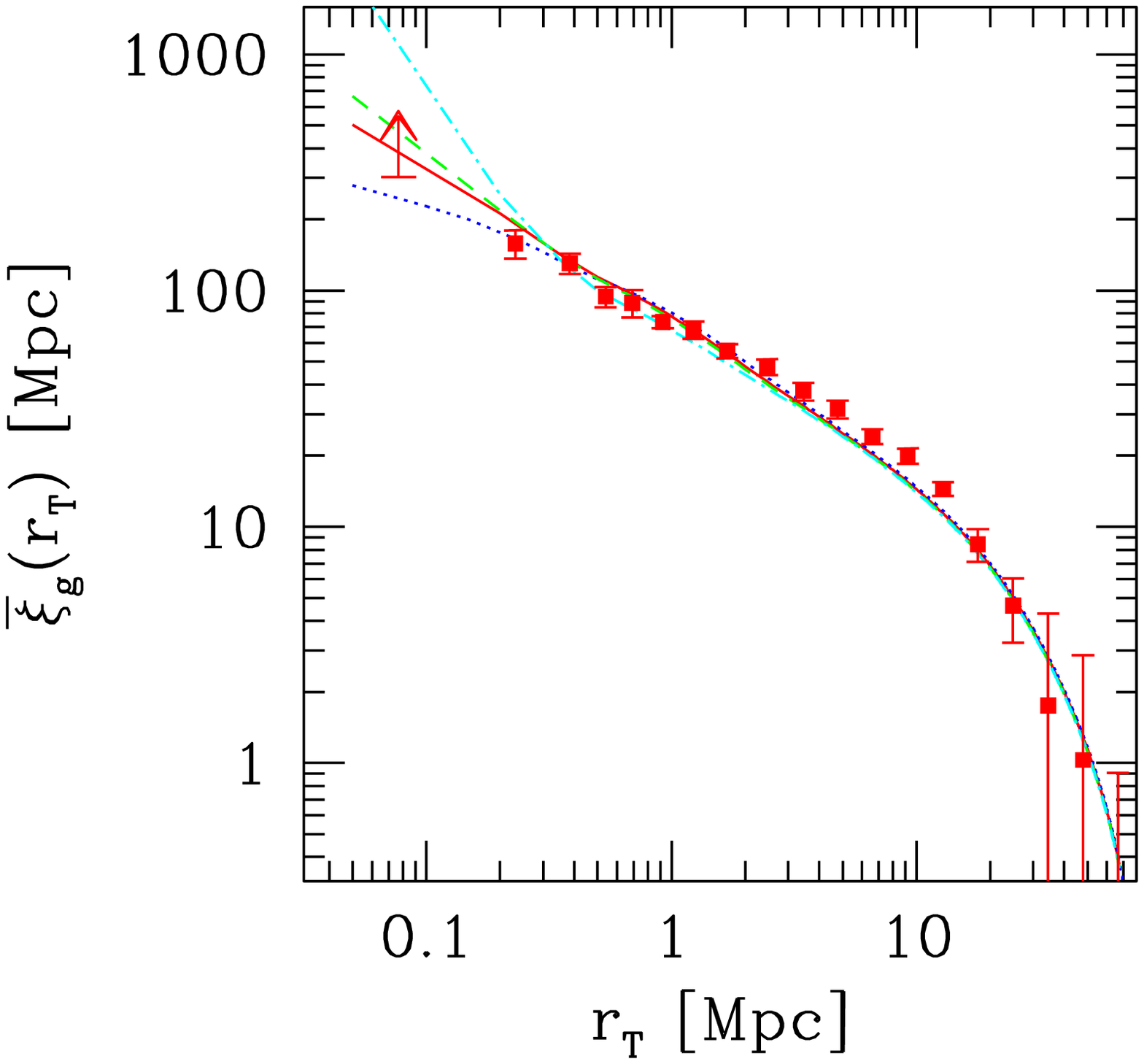}
\caption{Projected correlation function of late-type galaxies. Data-points 
represent the results from Madgwick et al. (2003), while solid, dashed, 
dotted and dashed-dotted curves illustrate the best-fit models respectively  
obtained for a NFW, a power-law with $\beta=2.5$, a power-law with $\beta=2$ 
and a power-law with $\beta=3$ galaxy distribution profiles and for a 
truncation radius $r_{\rm cut}=2\cdot r_{\rm vir}$. The arrow shows the lower limit 
to the measured correlation function set on small scales by fibre collisions in the 2dF 
survey (Ed Hawkins private communication).
\label{Fig:xilateall1}}
\end{figure}

A possible way out is therefore to deprive the intermediate-scale region of 
pairs of objects residing in different haloes. This can be done if one 
assumes that the distribution of late-type galaxies associated to a given
halo does not vanish at the virial radius, 
but extends to some larger distance. The mutual spatial exclusion of haloes 
then ensures that there will be no pairs of galaxies belonging 
to different haloes in the desired range.
Indeed, it is not surprising to find S0 and disc galaxies futher away
than a virial radius from
a cluster centre (see e.g. figure 8 in Dom\'inguez, Muriel \& Lambas, 2001).
We can attempt to let the galaxy distribution extend to -- say -- two 
virial radii.
As Figure (\ref{Fig:xilateall1}) shows, this assumption greatly reduces the 
discrepancy between models and measurements: the theoretical curves now have 
the right slope and amplitude on scales $0.5 \simlt r/[\rm Mpc]\simlt 2$ 
and the corresponding best fits to the data exhibit $\chi^2_{min}$ values 
which are (almost) as good as in the case of early-type galaxies (see Table~1).
Note that, even though we adopted a somehow ``ad hoc'' procedure to find a 
better description of the measurements, our finding seems to point out to 
the well-known phenomenon of {\it morphological segregation} (see e.g. 
Madgwick et al., 2003; Dom\'inguez, Muriel \& Lambas, 2001;
Giuricin et al., 2001 and Adami et al., 1998 for some 
recent results), whereby late-type galaxies tend to be found in the outer 
regions of groups and clusters, while early-type ones  
preferentially sink into the group or cluster centre.
A simple calculation in fact shows that, for scales $0.5 \simlt r/[\rm Mpc]
\simlt 2$, haloes which in our model are required to host star-forming 
galaxies up to two virial radii have 
masses in the range $10^{11.9}\simlt m/m_\odot\simlt 10^{13.7}$. As we will 
see better in the next Section, haloes within this mass range are 
expected to host on average 
$3\simlt \langle N_{\rm gal}\rangle\simlt 50$ late-type 
galaxies, limits which span from a small group to a cluster of galaxies.\\
Our result therefore does not contradict the well established fact that 
galaxies form within the virialized regions of dark matter haloes
(i.e. at a distance $< r_{\rm vir}$ from their centre). What it simply states 
is that -- possibly due to merging processes and accreting flows -- 
late-type galaxies in groups and clusters are found within 
their dark matter haloes up to distances from the centre corresponding to 
about two virial radii. Note that we are not claiming that these galaxies
are sling-shot towards their final position during the merging event.
The key idea is that, at a certain point, in the process of approaching a 
merging event, galaxies residing in the progenitors of a given halo will 
be associated with the final object itself as their host halos loose their 
identity by the formation of high density ``bridges'' which alter the output 
of cluster-finding algorithms like the friends-of-friends one (on which
both our mass function and bias parameters are based). 
This phenomenon might be less important for the 
population of early-type galaxies which, probably, form in galaxy
mergers that tend to be more concentrated within their host haloes. 

\begin{figure}
\vspace{8 cm} 
\includegraphics{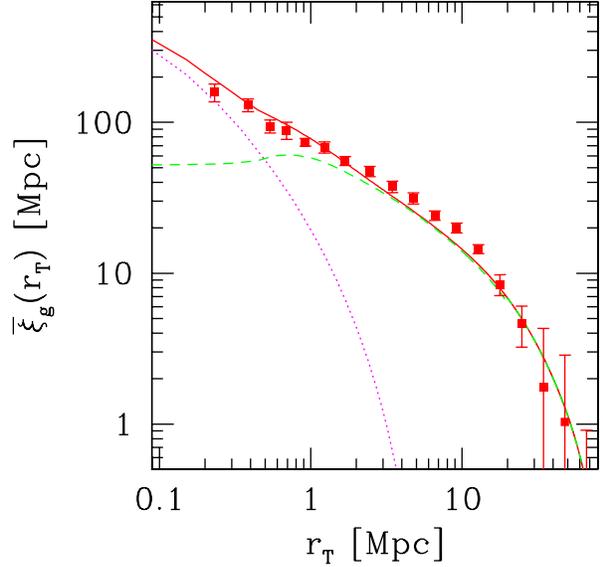}
\caption{Projected correlation function of late-type galaxies. Data-points 
represent the results from Madgwick et al. (2003), while the solid curve 
is the best fit to the measurements obtained for a halo number density of the 
form (10), with $\alpha_1=-0.4$, $\alpha_2=0.7$, $m_{\rm cut}=10^{11}m_\odot$,
$m_0=10^{11.4}m_\odot$ and for galaxies distributed within their dark matter 
haloes according to a NFW profile with $r_{\rm cut}=2\cdot r_{\rm vir}$. 
Dashed and dotted lines respectively indicate 
the contribution $\xi_g^{2h}$ from galaxies residing in different haloes and 
the $\xi_g^{1h}$ term originating from galaxies within the same halo.
\label{Fig:xilate}}
\end{figure}
 
We now discuss the
results on the halo occupation number and distribution profile obtained for 
the population of late-type galaxies under the assumption of a truncation 
radius corresponding to two virial radii. As Table~1 shows, the best fit to 
the data in this case is provided by a model with $\alpha_1\simeq -1$, 
$\alpha_2=0.7$, $m_{\rm cut}=10^{11.1}m_\odot$,
$m_0=10^{11.4}m_\odot$ and for galaxies distributed within their dark matter
haloes according to a PL2 profile, even though on the basis of this analysis we cannot really 
discard any $\rho(r)$ model but PL3 since they all show $\chi_{min}^2$ values 
within 1$\sigma$ from the favourite one (as for early-type galaxies, 
a model is
accepted as a fair description of the data if the corresponding 
$\chi^2_{min}$ lie within 
$\Delta\chi^2=3.53$ from the best fit).\\     

In order to shed some more light on the distribution profiles able to correctly 
describe the correlation function of star-forming galaxies, we have then adopted the same 
approach as in the previous sub-section and used the lowest-$r_{\rm T}$ ($\simeq 
0.07$~Mpc) measurement of $\xi_g(r)$ derived by the 2dF team for late-type objects 
(Ed Hawkins, private communication) 
as a lower limit to the true galaxy clustering strength, to be compared with the 
available models. This is done in Figure 5, where the lowest-scale data is represented 
by the vertical arrow. At variance with the case of early-type galaxies, 
the singular isothermal sphere model -- even though providing the best fit to the $r\simgt 
0.1$~Mpc data -- is now ruled out by the behaviour 
of the observed correlation function which in this regime needs steeper profiles. 
On the basis of the above comparison, one then has that only the PL2.5 a NFW density runs 
can still be accepted as reasonable descriptions of the data, with a slightly
stronger
preference given to the last model since it also provides the (second) best-fit to the 
observations on scales $r>0.1$~Mpc.\\
The projected galaxy correlation function originating from the 
combination of values given in Table 1 and the NFW profile is illustrated by 
the 
solid curve in Figure (\ref{Fig:xilate}). The dashed and dotted lines 
respectively indicate the contribution $\xi_g^{2h}$ from galaxies hosted
in different haloes and the $\xi_g^{1h}$ term given by galaxies 
residing within the same halo.
Note that, as in Figure (\ref{Fig:xiearly}), our best-fitting models tend
to underestimate the observed correlation at intermediate scales (between
5 and 15 Mpc) by $\sim 20$ per cent.

The uncertainties associated to the parameters describing the low-mass regime of
the halo occupation number seem to be bigger than what
found for passive galaxies. This result is however misleading, since the larger
errors quoted for $\alpha_1$ and associated to the 1$\sigma$ lower limit of
$m_{\rm cut}$ only reflect the fact that
the low-mass regime in equation (10) is practically non-existent
for star-forming galaxies as $m_0\simeq m_{\rm cut}$.
In this case one than has that the halo occupation number
behaves as the pure power-law:
$\langle N_{\rm gal}\rangle(m)=(m/m_0)^{\alpha_2}$ at almost all mass scales, 
where both $\alpha_2$ and $m_0$ are very well constrained.\\
As it was in the case of early-type galaxies, an analysis of the $\chi^2$ 
hypersurface helps understanding the interplay between 
the different parameters describing the halo occupation number.
We find that, while $\alpha_2$ and $m_0$ show little variability, 
$\alpha_1$ and $m_{\rm cut}$ are strongly covariant,
whereby higher values for the former quantity correspond to lower 
minimum masses associated to haloes able to host a star-forming galaxy.

As a final remark, we note that -- also for late-type 
galaxies -- values for $\alpha_1$, $\alpha_2$, $m_{\rm cut}$ and $m_0$ which 
provide the best fit to the data do not depend on the particular form 
adopted for the spatial distribution of galaxies within the haloes. 
Once again, this finding stresses the absence of degeneracy 
in the determination of quantities such as $\rho(r)$ and the average number of 
galaxies in haloes of specified mass, $\langle N_{\rm gal}\rangle(m)$: as long 
as one can rely on clustering measurements which probe the inner parts of  
dark matter haloes, both functional forms can be obtained with no degree of 
confusion from the same dataset.

\subsection{Scatter of the Halo Occupation Distribution 
and Number Density Profiles}
All our results for the galaxy density profiles are derived by assuming a 
specific functional form for $\langle N_{\rm gal} (N_{\rm gal}-1)\rangle(m)$, 
as 
given in equation (\ref{second}). In principle, this could bias our
determination of the spatial distribution of galaxies within a single halo  
(Berlind \& Weinberg 2002).
In order to understand the importance of this effect, we repeated the analysis
of the correlation functions
by assuming 3 different functional forms for the second moment of the
halo distribution function. In particular, we took $\alpha=1,0.3,0.1$ 
in equation (\ref{second}) independently of the halo mass. 
We found that
none of these models can fit the data as well as our original prescription.
However,
for purely Poissonian scatter, the best-fitting models for late-type galaxies 
are almost identical to our fiducial models, and the ranking of the 
density profiles does not change.
On the other hand, for early-type galaxies, Poissonian models are associated
with large values of $\chi^2$ (23.7 at best, for PL2) due to the
overabundance of power on small scales (the favorite values for $m_{\rm cut}$ 
and $m_0$ tipically are $\sim 10$ times smaller than in our fiducial case) but,
still, only the PL2 and NFW profiles are acceptable. 
When the scatter, instead, is strongly sub-Poissonian ($\alpha=0.1,0.3$)
we find that
it is practically impossible to get a good description of $\bar\xi_g$.
In fact, in this case,
the 1-halo term in the correlation function is heavily depressed
and one is forced to increase the value of $\alpha_2$ and $m_{\rm cut}$ 
to try to match the data. Anyway, 
all the models are unacceptable due to lack of power on small scales, and all 
the profiles are associated with nearly the same values of $\chi^2$.
In summary, even though we confirm the presence of some
degeneracy between the second moment
of the halo occupation distribution and the profile of the galaxy distribution 
as discussed in Berlind \& Weinberg (2002), we found that it is extremely hard
to find models that can give accurate description of the data. This means
that the apparent freedom in assuming a functional form for 
$\langle N_{\rm gal}(N_{\rm gal}-1)\rangle (m)$ 
is not such. As a consequence of this,
we are led to believe that our conclusions regarding the density profiles of 
2dF galaxies are indicative of a real trend,
even though they are indeed drawn in the framework of a specific model.

\section{The Galaxy Mass Function}

The results derived in the previous sections allow us to draw some conclusions 
on the intrinsic nature of galaxies of different types.
Figure (\ref{Fig:ng}) shows the average number of galaxies per dark matter 
halo of specified mass $m$ as obtained from the best fits to the clustering 
measurements of Madgwick et al. (2003). The dashed line describes the case 
for star-forming galaxies, while the solid line is for early-type objects.
 
Late-type galaxies are found in haloes with masses greater than $\sim 10^{11}\;
m_\odot$ (even though this figure, especially in its lower limit, is not determined with 
a great accuracy given the interplay between $m_{\rm cut}$ and $\alpha_1$), 
and their number increases with the mass of the halo which hosts
them according to a power-law (except in the limited region $10^{11.0}
\le m/m_\odot < 10^{11.4}$) with slope $\alpha_2=0.7$ and
normalization $1/m_0=10^{-11.4} m_{\odot}^{-1}$.\\
Early-type galaxies instead start appearing within haloes of noticeably higher masses, 
$m\simgt 10^{12.6}\;m_\odot$. In the low-mass region (i.e. for $m\simlt 10^{13.5}\;
m_\odot$), the data seems to indicate that each halo is on average populated by 
approximately one passive galaxy, even though results in this mass range are affected by 
some uncertainties. More solid are 
the findings in the high mass ($m\simgt 10^{13.5}\; m_\odot$) regime, which 
show 
$\langle N_{\rm gal}\rangle$ to increase with halo mass as a power-law of 
slope 
$\alpha_2\simeq 1.1$, steeper than what found for the population of late-type 
galaxies. One then has that the average number of star-forming galaxies
within a halo does not increase with its mass as fast as it happens
for passive galaxies, even though late-type objects are found to dominate the
2dF counts at all mass scales. 
This result seems in disagreement with the observational
evidence that early-type galaxies are preferentially found in
clusters, while star-forming galaxies mainly reside in relatively underdense
regions. The discrepancy is however
only apparent since, while in the Madgwick et al. (2003) analysis the
population of early-type galaxies is an homogeneous sample, the class of
late-type galaxies includes any object which has some hint of star-formation
activity in its spectrum (see Section 3). This implies that anything --
from highly irregular
and star-bursting galaxies to S0 types -- will be part of the star-forming
sample. The lack of a characteristic mass for the transition from haloes
mainly populated by late-type galaxies to haloes principally inhabited
by passive objects then finds a natural explanation if one considers that S0
galaxies -- also preferentially found in groups and clusters -- are in
this case associated to the population of late-type objects, making up for 
about 35 per cent of the sample (Madgwick et al., 2002).
We also remind that both the 2dF and its parent APM surveys select 
sources in the
blue ($b_j$) band, therefore creating a bias in favour of star-forming galaxies
which are more visible than passively evolving ellipticals in this wavelength 
range.

It is interesting to compare our results with those by
van den Bosch et al. (2003) who used 2dF data to estimate the conditional
luminosity functions of early- and late-type galaxies. They combined their 
results to obtain the mean halo occupation number of galaxies in given 
absolute-magnitude ranges. Since our analysis is based on a apparent-magnitude
limited sample, this complicates the comparison.
For their faintest late-type galaxies, the behaviour of 
$\langle N_{\rm gal}\rangle(m)$ is relatively similar to our findings 
with a cutoff
below $\sim 10^{11} m_\odot$, a small decrement up to $10^{11.5} m_\odot$
and a power-law regime $d\log \langle N_{\rm gal}
\rangle(m)/d\log m\simeq 0.6$ for larger masses. 
As for early-type galaxies, results are in good agreement, with
a power-law high-mass regime with slope $d\log \langle N_{\rm gal}
\rangle(m)/d\log m\simeq 1$. 
However, van den Bosch et al. (2003) find a less sharp 
cutoff at small masses,
with $\log (m_0/m_\odot)\sim 12.5$ and $\langle N_{\rm gal}(m)\rangle$ gently
declining with decreasing $m$ till to $10^{10.5-11} m_\odot$. 
Given all the systematic uncertainties belonging to the two method of analysis
the similarity of the results is remarkable, confirming the potential of the
halo occupation distribution method.

\begin{figure}
\vspace{8 cm} \includegraphics{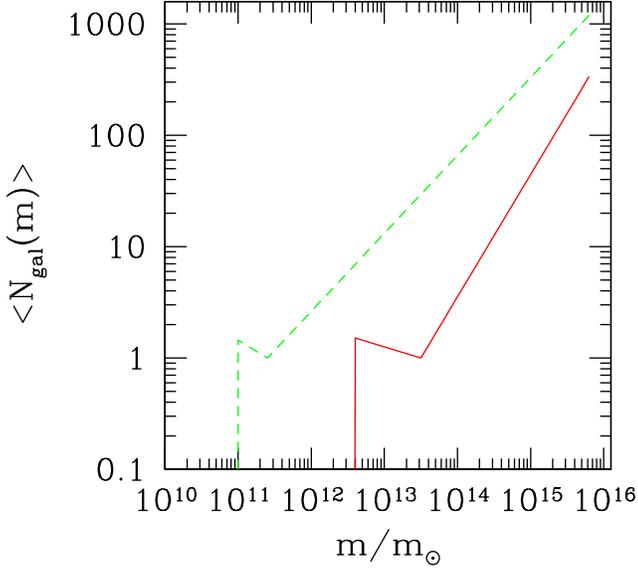}
\caption{Average number of galaxies $\langle N_{\rm gal}(m)\rangle$ per dark matter halo 
of specified mass 
$m$ (expressed in $m_\odot$ units). The solid line represents the case for 
early-type galaxies, while the dashed line is for late-type objects. 
\label{Fig:ng}}
\end{figure} 

\begin{figure}
\vspace{8 cm} \includegraphics{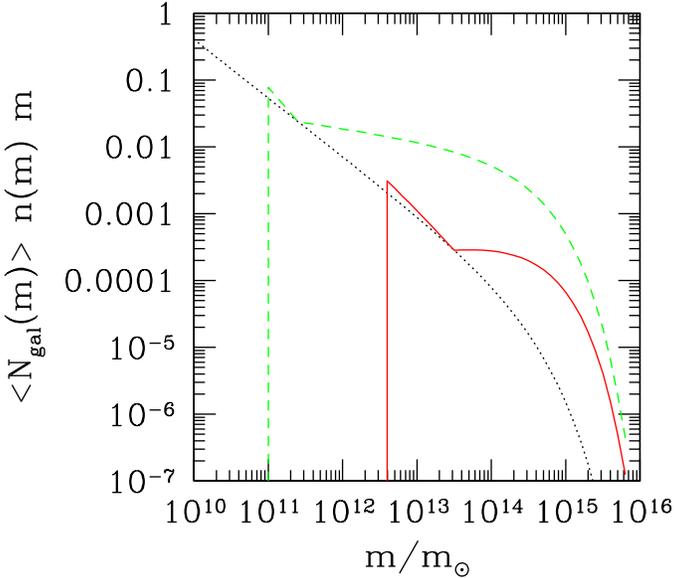}
\caption{Number of galaxies per unit of (log) dark matter mass and volume
(in ${\rm Mpc}^{-3}$ units). 
The solid line represents the result for 
early-type galaxies, while the dashed line is for late-type objects. 
For comparison, the dotted line indicates the Sheth \& Tormen (1999) mass 
function $n(m)$ of dark matter haloes.
\label{Fig:mf}}
\end{figure}

Finally, Figure (\ref{Fig:mf}) shows the ``galaxy mass function'' i.e. the 
number density of galaxies per unit of dark matter mass and volume as 
obtained by 
multiplying the average number of galaxies found in a halo of specified mass 
$\langle N_{\rm gal}(m)\rangle$ by the halo mass function (\ref{eq:nm}). 
Again, the solid 
curve identifies the case for early-type galaxies, while the dashed one is 
derived for the population of star-forming galaxies. The dotted line 
indicates the Sheth \& Tormen (1999) halo mass function (\ref{eq:nm}). 

\section{Conclusions}
Results from Madgwick et al. (2002) and (2003) on the correlation function 
and luminosity function of $\sim 96,000$ 2dFGRS 
galaxies with $0.01 <z < 0.15$
have been used to investigate some of the properties of early- and 
late-type galaxies, such as the so-called halo occupation number 
$\langle N_{\rm gal}\rangle$ (i.e. 
the mean number of sources that populate a halo of given mass $m$) and 
the spatial distribution of such galaxies within their dark matter haloes.

In order to perform our analysis, we have considered four distribution 
profiles: three power-laws of the form $\rho(r)\propto r^{-\beta}$ with 
$\beta=2,2.5,3$ and a NFW profile chosen to mimic the assumption for 
galaxies within haloes to trace the distribution of dark matter. As a first 
approximation, all the profiles have been truncated at the halo virial 
radius.\\
For consistency with results from semi-analytical models (see e.g. Benson et 
al., 2001), the halo occupation number was parametrized by a broken power-law 
of the form $(m/m_0)^{\alpha_1}$ in the low mass regime 
$m_{\rm cut}\le m\le m_0$ and $(m/m_0)^{\alpha_2}$ at higher masses, where 
$m_{\rm cut}$ is the minimum mass of a halo that can host a galaxy.
    
The resulting theoretical average number density $\bar{n}_g$ and 
galaxy-galaxy correlation function -- sum of 
the two terms $\xi_g^{1h}$, representing the contribution from galaxies 
residing within the same halo, and $\xi_g^{2h}$ which considers pairs 
belonging to different haloes -- have then been compared with the 
observations in order to determine those models which provide the best 
description of the data both 
in the case of late-type and early-type galaxies. Note that, at variance with 
previous works which only considered a linear $\xi_g^{2h}$, 
our analysis provides a full treatment for non-linearity and also includes  
the assumption of halo-halo spatial exclusion, in a way that makes the model 
entirely self-consistent.  

The main conclusions are as follows:
\begin{enumerate}
\item 
Early-type galaxies are well described by a halo occupation number of the 
form broken power-law (10) with $\alpha_1\simeq -0.2$, $\alpha_2\simeq 1.1$, 
$m_{\rm cut}\simeq 10^{12.6}m_\odot$ and $m_0\simeq 10^{13.5}m_\odot$, where the two 
quantities which determine the intermediate-to-high mass behaviour of 
$\langle N_{\rm gal}\rangle$ are measured with a good accuracy.
\item 
No model can provide a reasonable fit to the correlation function 
of late-type galaxies since they all show an excess of power with respect 
to the data on scales $0.5\simlt r/[\rm Mpc]\simlt 2$. In order to obtain 
an acceptable description of the observations, one has to assume that 
star-forming galaxies are distributed within haloes of masses comparable 
to those of groups and clusters up to two virial radii. This result  
is consistent with the phenomenon of morphological segregation whereby 
late-type galaxies are mostly found in the outer regions of groups or 
clusters (extending well beyond their virial 
radii),  while passive objects preferentially sink into their centres.
\item With the above result in mind, one finds that late-type galaxies 
can be described by a halo occupation number of the 
form single power-law with $\alpha_2\simeq 0.7$, 
$m_{\rm cut}\simeq 10^{11}m_\odot$ and $m_0\simeq 10^{11.4}m_\odot$, where the 
quantities which describe $\langle N_{\rm gal}\rangle$ in the high-mass regime 
are determined with a high degree of accuracy.
\item 
Within the framework of our models,
galaxies of any kind seem to follow the underlying distribution of dark 
matter
within haloes as they present the same degree of spatial concentration.
In fact the data indicates both early-type and late-type galaxies to be 
distributed 
within their host haloes according to NFW profiles.
We note however that, even though early-type galaxies can also be described by
means of a shallower distribution of the form $\rho(r)\propto
r^{-2}$, this cannot be accepted as a fair modelling of the data in the case
of late-type galaxies which instead allow for somehow steeper ($\beta\simeq 2.5)$
profiles. In no case a $\beta=3$ density run can provide an acceptable description of the observed correlation function.
These conclusions depend somehow on assuming a specific functional form
for the second moment of the halo occupation distribution. We showed,
however, that there is not much freedom in the choice of this function
is one wants to accurately match the observational data. 
\end{enumerate}

An interesting point to note is that results on the spatial distribution 
of galaxies within haloes and on their halo occupation number are 
independent from each other. There is no degeneracy in the determination of 
$\langle N_{\rm gal}\rangle$ and $\rho(r)$ as they dominate the behaviour 
of the two-point correlation function $\xi_g$ at different scales. Different 
distribution profiles in fact principally determine the slope of 
$\xi_g$ on small enough ($r\simlt 1$~Mpc) scales which probe the inner 
regions of the haloes, while the halo occupation number is mainly responsible 
for the overall normalization of $\xi_g$ and for its slope on large-to-intermediate 
scales. 

Our analysis shows that late-type galaxies can be hosted in haloes with masses 
smaller than it is the case for early-type objects. This is probably due 
to the fact that early-type galaxies are on average more massive (where 
the term here refers to stellar mass) than star-forming objects, especially 
if one considers the population of irregulars, and points to a relationship  
between stellar mass of galaxies and mass of the dark matter haloes which host 
them.

The population of star-forming galaxies is found to be the dominant one
at all mass scales, result that can be reconciliated with the well established
observational fact that early-type galaxies are preferentially found in
clusters, while star-forming galaxies mainly reside in relatively underdense
regions, by considering that about a third of the class of late-type sources
in our sample is made of S0 galaxies, which are also preferentially found in groups
and clusters. We also stress that both the 2dF and
its parent APM surveys select sources in the
blue band, therefore creating a bias in favour of star-forming galaxies
which are more visible than passively evolving ellipticals in this wavelength range.
 
As a final remark, we note that the results of this work are partially 
biased by the need to assume a pre-defined functional form for both 
the halo occupation number and the variance about this quantity. 
High precision measurements of higher moments of the galaxy distribution 
function (such e.g. the skewness and the kurtosis) are of crucial 
importance if one wants to determine the distribution of galaxies within 
dark matter haloes in a non-parametric way, i.e. without the necessity to 
rely on any ``a priori'' assumption.
In the near future, results from the 2dF and SSDS galaxy redshift surveys 
should be able to fill this gap.

\noindent
\section*{ACKNOWLEDGMENTS}
MM thanks the Institute of Astronomy for the warm hospitality during visits 
which allowed the completion of this work. 
CP has been partially supported by the Zwicky Prize Fellowship program at
ETH-Z\"urich and by the European Research and Training Network
``The Physics of the Intergalactic Medium''.
We are extremely grateful to Darren 
Madgwick, Ed Hawkins and Peder Norberg for endless discussions on the 2dF data
and results and to Ofer Lahav for many clarifying conversations.

\end{document}